\documentstyle[11pt,paspconf,epsf,itzidef]{article}

\begin{document}

\title{When SN explode in a starburst: trasitions among AGN types}
\author{Itziar Aretxaga}
\affil{Instituto Nacional de Astrof\'{\i}sica, \'Optica y Electr\'onica,
Aptdo. Postal 51 y 216, Puebla, Pue., Mexico}

\def\Ha{\mbox{H{\small $\alpha$}}}
\def\Hb{\mbox{H{\small $\beta$}}}
\def\ewHa{\hbox{$W_{\mbox{\scriptsize H{\tiny $\alpha$}}}$}}
\def\ewHb{\hbox{$W_{\mbox{\scriptsize H{\tiny $\beta$}}}$}}
\def\Ho50{\hbox{H$_{\mbox{\scriptsize 0}} = 50$~\uniHo \ }}
\def\uniHo{\hbox{Km s$^{-1}$ Mpc$^{-1}$}}
\def\uniden{\hbox{cm$^{-3}$}}
\def\SNrate{\hbox{$\nu_{\mbox{\scriptsize SN}}$}}
\def\LBstar{\hbox{$L_{\mbox{\scriptsize B}}^{\star}$}}
\def\LBsun{\hbox{L$_{{\mbox{\scriptsize B}}}^{\odot}$}}
\def\lsim{\mathrel{\lower2.5pt\vbox{\lineskip=0pt\baselineskip=0pt
           \hbox{$<$}\hbox{$\sim$}}}}
\def\gsim{\mathrel{\lower2.5pt\vbox{\lineskip=0pt\baselineskip=0pt
           \hbox{$>$}\hbox{$\sim$}}}}
\def\tsg{\hbox{$t_{\mbox{\scriptsize sg}}$}}

\begin{abstract}
Among the Active Galactic Nuclei (AGN) that have suffered type mutations, 
the classically classified Seyfert~2 
NGC~7582 recently found at a Seyfert~1 stage, is perhaps one of the most 
puzzling cases, since it defies a well supported case for unification. 
We investigate the possibility that this type transition,
common to other 12 AGN, is driven by supernovae (SN) 
exploding in a nuclear/circumnuclear starburst.
\end{abstract}


\section{Introduction}
  Variability is a common characteristic of AGN.
  Most variability studies
  concentrate their attention on luminous Seyfert~1 nuclei and QSOs, which are
  known to develop prominent variations.
  But even among brands which
  classically have been regarded as quiescent, LINERs and Seyfert~2 nuclei,
  there is an increasing number of reported cases of variations in both
  development of broad lines
  and increase of continuum luminosity. 

\section{Seyfert~1 mutation of the classical Seyfert~2 NGC~7582}

  The recent discovery that NGC~7582 has mutated into a type~1 Seyfert$^{1,2}$
is perhaps one of the most puzzling cases. This classical Seyfert~2
shows phenomenology that has been interpreted in support of unified
schemes of AGN$^3$: a sharp-edged [O~III] outflow in the form of a cone is 
detected$^4$; optical spectropolarimetry does not reveal any hidden broad
line region (BLR), but since the
far-IR colours $60\mu\mbox{m} - 25\mu$m are very red, the absence has been 
taken in support for an edge-on thick torus 
able to block even  the light scattered towards the 
observer$^5$; indeed
a large column density of neutral H also blocks the hard X-rays, 
implying a very large obscuration$^6$.
The presence of stars in the nucleus is now firmly established. There is
a steep gradient of \Ha\  perpendicular to the [O~III] cone,
which reveals a
1~kpc disk of H~II regions oriented at $60^o$ from the plane of the galaxy$^4$.
The CO absorption lines and large near-IR light-to-mass ratio are
similar to those of H~II galaxies and a factor of 5 larger 
than those of normal galaxies, indicating that red supergiants dominate
the light of the inner 200~pc at those wavelengths$^7$.

Figure~I shows spectra of the nucleus of 
NGC~7582 as seen on January 6th 1994$^8$, the long known Seyfert~2 stage
that lasted until 1998 June 20th$^9$,
and on July 11th, when it was discovered to have transited to a
Seyfert~1 stage$^1$. 

\begin{figure*}
    \cidfig{5in}{26}{150}{590}{340}{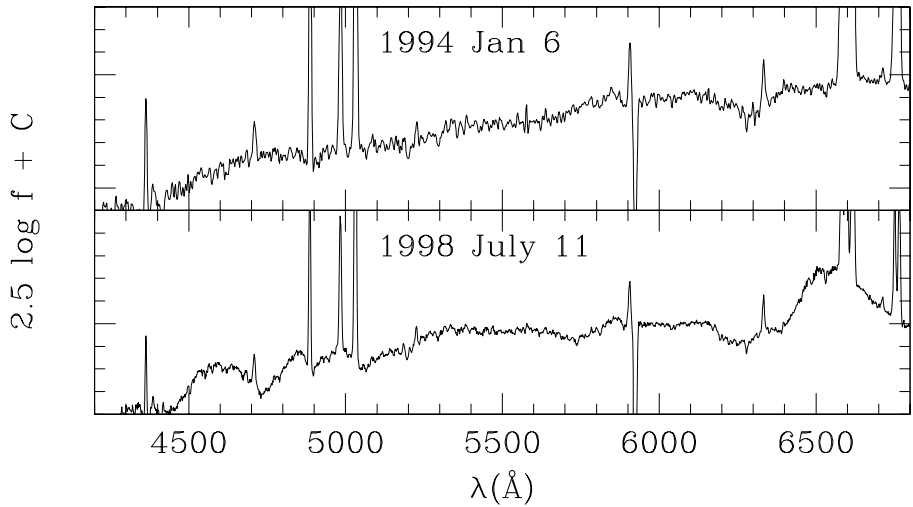}
    \caption{2~arcsec extractions of spectra of the nucleus of NGC~7582}
\end{figure*}

This type transition,
if rare, is not unprecedented. Three other classically classified
Seyfert~2 nuclei and one 
{\sc liner} had suddenly developed broad lines in the past
(Mrk~6, Mrk~993 and Mrk~1018 and NGC~1097), and 
seven Seyfert~1 and a radio-galaxy lost their broad components and
become Seyfert~2 (NGC~1566, NGC~3516, NGC~4151, NGC~5548, NGC~6814, 
NGC~7603, Mrk~372 and 3C~390.3).

Some of these transitions have been interpreted in the framework of
capture and disruption of stars by a supermassive black hole$^{10}$
or by a variable reddening in the dusty torus that hides
the BLR in Seyfert~2$^{11}$. 
In the case of NGC~7582 these two interpretations
conflict with the high extinctions measured for the thick dusty torus
($A_V\approx 230$~mag), that would block all the optical light coming from
the surroundings of the black hole$^{2}$. We can thus say, that the
mutation experienced by NGC~7582 defies the framework postulated
by unified schemes.

  If the idea of a torus is to survive, 
the spectral change must be attributed 
to processes that occur around the torus, and which are not necessarily
related to the 
central engine of the AGN. An attractive alternative is to attribute this
Seyfert~1 activity to type~IIn SN going off in
the nuclear starburst detected at a radius $r<100$~pc $^7$.

\section{Type IIn supernovae: SN~1988Z}

The spectra of SN~IIn are characterized by the 
presence of prominent narrow emission lines (hence the 'n')
sitting on top of broad
components of up to FWHM$\approx 20000$~km/s which look very similar to
those of Seyfert~1 nuclei and QSOs$^{12}$, and 
don't show the characteristic broad P-Cygni signatures of standard SN.
SN~IIn are normally associated with regions of star-formation.

\begin{figure*}
    \cidfig{5in}{26}{150}{590}{430}{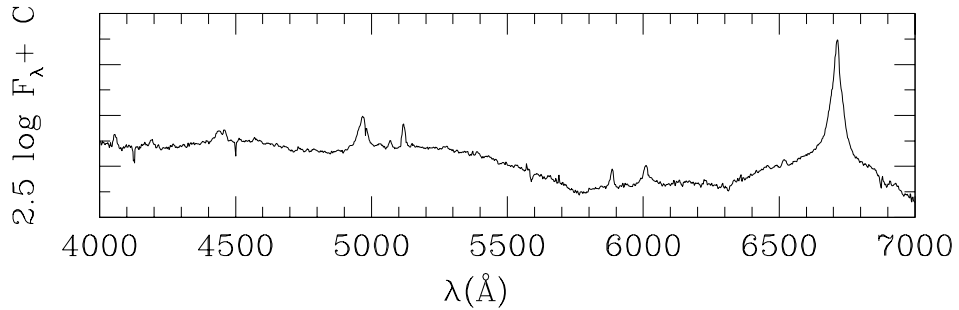}
    \caption{Spectrum of SN~1988Z at age 115~d from discovery} 
\end{figure*}

Figure~II shows a spectrum of SN~1988Z, the best followed-up SN of this type,
near maximum$^{13}$.
This  is an exceptionally bright and peculiar SN in its
spectro-photometric
properties:\\
	\hspace*{0.5cm} $\bullet$ 
It is characterized by an extremely slow decay of
luminosity 
after maximum light, which makes it at day 600 approximately 5~mag
brighter in $V$-band than standard type II SN$^{14}$.\\
	\hspace*{0.5cm} $\bullet$  It has a strong  \Ha\ emission, with 
peak luminosities of about $4 \times 10^{41}$~erg~s$^{-1}$ 
(for \Ho50)
at day 200$^{13}$. This
prodigious luminosity is 5 orders of magnitude larger than that
of SN~1987A.\\
	\hspace*{0.5cm} $\bullet$  Very high-ionization coronal lines 
(e.g. [Fe X]\ldo{6375},
[Fe XI]\ldo{7889-7892})
are identified in the optical spectra, at least until day 492$^{13}$.\\
	\hspace*{0.5cm} $\bullet$ 
At 2 to 20cm  it is one of the most powerful 
radio-SN in the sky,
with peak-luminosities up to 3000 times that of remnants like Cas~A$^{15}$\\
	\hspace*{0.5cm} $\bullet$ 
	Even 6~yr after maximum the SN shows a hard X-ray
emission$^{16}$ of more than $10^{41}$~erg~s$^{-1}$ \\
	\hspace*{0.5cm} $\bullet$  The total integrated energy radiated by
this event in 8.5~yr of evolution is at least $2 \times 10^{51}$~erg,
 and probably close to $10^{52}$~erg, in contrast with the canonical 
$10^{49}$~erg of standard SN$^{17}$.\\

Figure~III shows a combined representation of the above properties.
From this we derive that the output of energy is dominated by optical
to X-ray emission, with a spectral energy distribution which
strongly departures from the classical black-body fit of typical SN$^{17}$.

\begin{figure*}
   \cidfig{5in}{20}{144}{580}{710}{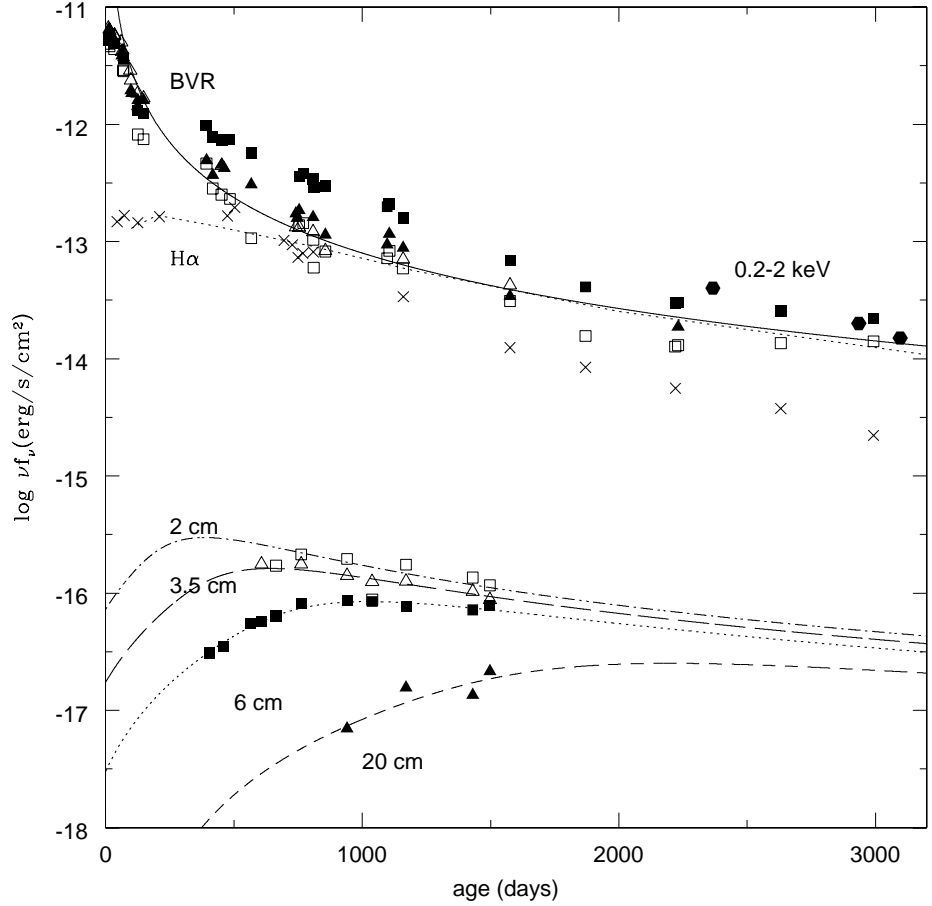}
   \caption{Evolution of the $\nu f_\nu$ light curve in radio to X-ray
bands. In the upper part of the panel open  triangles correspond to $B$-band,
solid triangles to $V$-band, solid 
squares to $R$-band, empty squares to $Rc$ band,  solid
hexagones to the ROSAT 0.2 to 2 keV band and crosses to \Ha. The solid and 
dashed lines are models for the bolometric light and \Ha\ evolution of a 
cSNR.
In the lower part of the panel, VLA radio data at 2, 5.5, 6 and 20cm is 
represented with empty squares, empty triangles, solid squares and empty 
triangles. The 
line fittings correspond to the models of Van Dyk et al$^{15}$.
}
\end{figure*}

These properties can be interpreted in the light of
quick reprocessing of the kinetical energy released in the explosion 
by a dense circumstellar medium (CSM)$^{17, 18}$,
and thus explain the phenomenon as a young and compact supernova remnant 
(cSNR) rather than as a SN. Radiative cooling is expected to become
important well before the thermalization of the ejecta is complete.
As a result, the shocked material undergoes a rapid condensation
behind both the leading and reverse shocks. These high-density thin
shells, the freely expanding ejecta and the still unperturbed interstellar
gas are all ionized by the radiation produced in the shocks, and are
responsible for the complex emission line structure observed in 
these objects.

In the case of SN~1988Z, a direct measurement
of the CSM density was possible$^{14}$.
The value determined from 
the [OIII]\ldo{4363}\ to [OIII]\ldo{5007}\
forbidden line ratio in the early stages of the evolution
is between  $4 \times 10^6$ and
$1.6 \times 10^7$~\uniden.

The semi-analytical behaviour of a cSNR$^{18}$ in comparison with the data
of SN~1988Z is shown with lines in figures III and VI.

\section{Starbursts that explode type IIn SN}

\begin{figure*}
    \cidfig{5in}{40}{150}{585}{330}{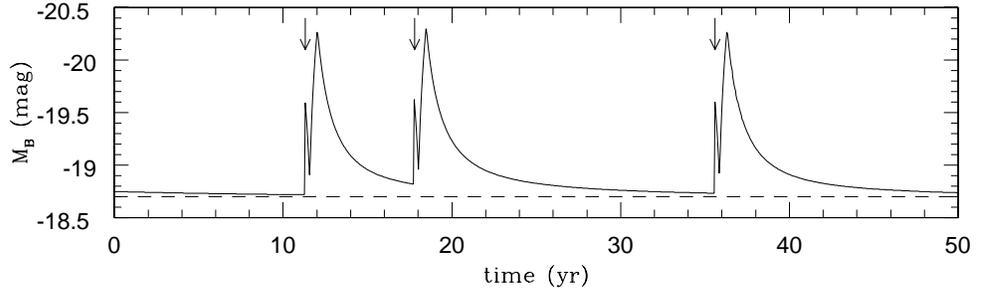}
    \caption{Theoretical light curve of a massive stellar cluster
undergoing a SN rate $\SNrate \approx 0.1$ yr$^{-1}$. The arrows in the
diagram represent
the moments in which the SN explodes and the dashed lines the luminosity
level coming from main sequence stars$^{21}$.}
\end{figure*}

There is little doubt that if a SN~IIn goes off in the center of a normal 
galaxy, the nucleus would be classified as a 
Seyfert~1  while 
the prominent broad lines remain visible. In fact, there has been
a succession of theoretical works that explain the phenomenology 
of lines and continuum at UV to near-IR 
wavelengths in Seyfert~1 nuclei in terms of a starburst that undergoes
SN~IIn explosions (see Terlevich R.J. in this volume).
From an age 10--60~Myr, 
a starburst sustains a SN~rate directly proportional 
to the blue light emitted by the stars in the cluster,
with a value that is almost independent on the initial mass function$^{19}$:
\begin{equation}
 \SNrate / \LBstar \approx 2 \times 10^{-11} \mbox{\ \ yr}^{-1}
 \mbox{\LBsun}^{-1} \mbox{\ \ \ .}
 \end{equation}

If we take the light evolution of SN~1988Z as typical of the
SN exploding in a cluster (but beware that SN~IIn show
some heterogeneity in their light curves$^{20}$), we can construct 
some models to predict  type transitions generated by SN~IIn$^{21}$.
Figure~IV shows a Monte Carlo simulation of the light curve of a cluster of 
$\MBstar \approx-18.7$ (marked with a dashed line). The SN light 
evolution has been represented
by the semi-analitical fit shown in figure~III, and each 
SN explosion is marked with an arrow.

According to the photoionization models for cSNR$^{18}$, 
the equivalent widths of \Ha\ and \Hb\ change in
time from
$\ewHa / \SNrate \approx 37.7 $~\hbox{\AA\ yr},
$\ewHb / \SNrate \approx 6.1  $~\hbox{\AA\ yr}
for 4\tsg\ to
$\ewHa / \SNrate \approx 13.1 $~\hbox{\AA\ yr},
$\ewHb / \SNrate \approx 1.4 $~\hbox{\AA\ yr}
for 8\tsg, where \tsg\ is the characteristic time of evolution of the
cSNR. If we adopt 20~\AA\ as the observable limit below which an object
is classified as a Seyfert~1.9 (in case $\ewHb \lsim 20$ \AA\ but
$\ewHa \gsim 20$ \AA) or a Seyfert~2 (in case $\ewHa, \ewHb \lsim
20$~\AA), the transitions take place when the total light emitted by the
cluster (solid line in figure~IV) is less than about 0.14 and 0.01~mag above
the stellar level (dashed line) respectively. 
The activity level is
recovered once the light curve crosses those limits in the opposite
direction due to a new cSNR.

We can then estimate the time spent by
these clusters as Seyfert nuclei of types~1.9 or 2$^{22}$. 
The values obtained are just upper limits to the time spent in 
quiescent stages,
since our approach ignores the secondary pulses that occur in the evolution
of cSNR due to cooling instabilities.
The less luminous systems are the ones that experience longer quiescent
stages. 
In low-luminosity systems ($M_B > -22.5$~mag)
 the low SN rates derived
($\SNrate \lsim 1$~yr$^{-1}$) gives non negligible time
scales for states
in which no cSNR could contribute 
to the existence of broad lines in the spectrum, and 
thus
transitions between Seyfert types 1 and 2 can be possible.
Figure~V compares  the luminosity of the transient AGN 
as compared with that theoretical limit, which lies
above the measured luminosities.

\begin{figure*}
    \cidfig{5in}{40}{150}{585}{374}{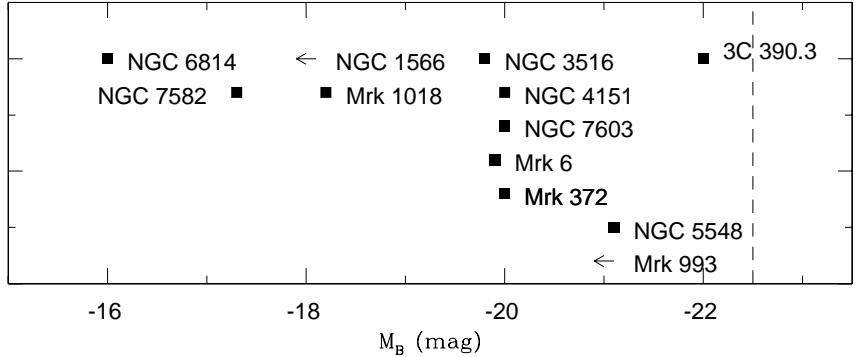}
    \caption{Number of transient AGN per luminosity interval. 
Squares represent mean luminosities, and 
arrows indicate upper limits to these quantities.
The luminosity threshold for allowed transitions is marked with a
line}
\end{figure*}

\section {a type IIn SN in the nucleus of NGC~7582?}
  The luminosity of the inner 3~arcsec of NGC~7582 before the transition
is $V \approx 15$~mag. The luminosity profile
at this stage 
is not very peaked, as shown in HST images,
where the luminosity of the nucleus changes by 3~mag when using
apertures ranging from 0.5 to 0.1~arcsec. 
The stellar populations of the nucleus at optical wavelengths 
are best represented by old 
ages ($t>100$~Myr) contaminated in a 12\%\ by a reddened 
$E_{B-V}\approx 0.6$~mag starburst$^{23}$.
Therefore, the intrinsic luminosity of the starburst nucleus is about 
$M_B\approx -17.3$~mag and
the SN rate must then be 
$\SNrate \approx0.02$~yr$^{-1}$. 
If all the SN explosions generate SN~IIn, there 
is a 33\% probability of having detected
a transition to a Seyfert~1 stage in the last 30~yr.

In order to investigate this possibility for the origin of the broad lines,
profile and light evolution might be of interest.
Figure~VI plots the evolution of the 
\Ha\ line width and luminosity of NGC~7582 
as a function of time compared with
the values of SN~1988Z$^{17}$  and the Seyfert~1 NGC~5548$^{24}$. 
The data of NGC~7582 was obtained from various ESO telescopes$^{2}$ 
in La Silla, Chile, 
and from the 2.1m in Cananea, Mexico (previously unpublished 
data). The spectra were internally flux-calibrated to the same relative scale 
using [N~II]\ldo{6583} and, independently, also [O~III]\ldo{5007}.
The zeros in the time axis of figure~VI 
for the comparison are uncertain. We have opted to
match the light curves of SN~1988Z, NGC~7582 and NGC~5548 at the maximum in 
\Ha\ light. This also matches the minimum 
in the light evolution of NGC~5548 as the onset of the
'elementary unit of variation' in Seyfert~1 nuclei$^{25}$.
All fluxes have been normalized to the maximum value. 
The line-width evolution of NGC~7582 closely resembles the early evolution
of SN~1988Z, if the similarity of flux evolution is not clear due to the
scarcity of the data. At these early stages, large deviations from the
semi-analytical solution of the evolution of a cSNR, represented as a 
dashed line, are expected due to the formation and collision of the 
shell structures generated by the outer and reverse shocks as they sweep 
circumstellar and ejected material$^{18}$. However,
these light oscillations predicted by theory are as yet unchecked in
real SN~IIn. These oscillations are however 
clearly present in the behaviour 
of the light evolution of typical Seyfert~1 nuclei, as NGC~5548 
represented with dotted lines. If the event seen
in NGC~7582 is a cSNR in the nuclear starburst region, a behaviour
similar to SN~1988Z and the peak of NGC~5548 
represented in Figure~VI is expected. 

A close 
multi-wavelength monitoring of this object is thus required if we want 
to elucidate the mechanism that has created the broad-lines despite the
supposedly still existing thick torus that blocks 
the inner nuclear region from the line of sight.

\begin{figure*}
    \cidfig{5in}{40}{150}{585}{430}{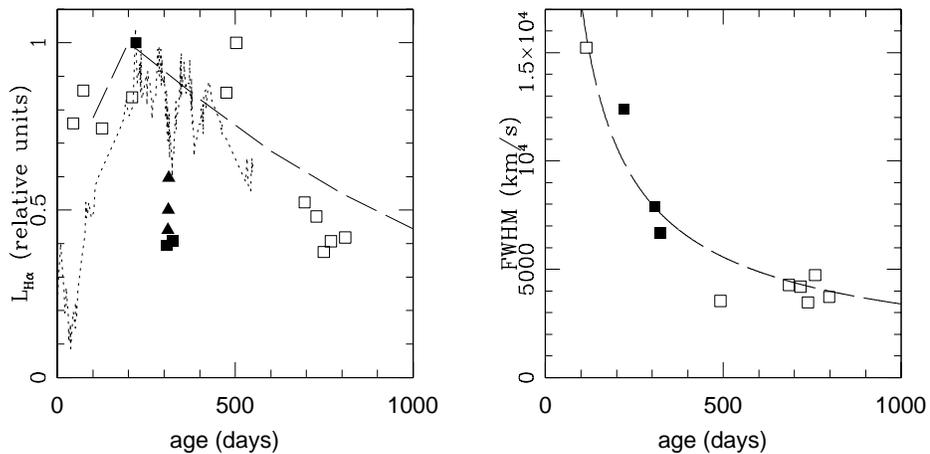}
    \caption{Evolution of \Ha\ line with and luminosity for NGC~7582
(filled squares and triangles), 
SN~1988Z (empty squares), NGC~5548 (dotted line)
and a cSNR model (dashed line)}
\end{figure*}

\acknowledgments
I would like to acknowledge my collaborators in the several projects that have
been joined to shape this paper: S. Benetti, B. Joguet, D. Kunth, 
J. Melnick, A. Fabian, R.J. Terlevich and J.R. Valdes.
Especial thanks go to B. Joguet and J.R. Valdes who were crucial in providing
me with the data of NGC~7582, keeping the flow open until hours before 
my talk.

\end{document}